\documentclass[a4paper,11pt]{article}
\pdfoutput=1 

\usepackage{jheppub} 

\usepackage[T1]{fontenc} 
\usepackage{subfigure}
\usepackage{relsize}
\usepackage{array,multirow}
\usepackage{soul}
\usepackage{secdot}
\usepackage{subfigure}
\usepackage{dsfont}
\usepackage{hyperref}
\usepackage{txfonts}
\usepackage{newlfont}
\usepackage{times}
\usepackage[utf8]{inputenc}

\title{\boldmath Quantum Gravity Effect on The Time History of The Universe}

\author[a]{I. A. Elmashad}
\author[a]{Asmaa G. Shalaby}

\affiliation[a]{Physics Department, Faculty of Science, Benha University, Benha 13518, Egypt.}

\emailAdd{ibrahim.elmashad@fsc.bu.edu.eg}
\emailAdd{asmaa.shalaby@fsc.bu.edu.eg}

\abstract{The QCD phase transition exhibits different periods during the evolution time of the universe. We explore
the cosmological implications of QCD phase transition in the early universe, which was mainly full of QGP matter, through the Friedmann equations. This is performed by studying the effect of GUP on the MIT bag model. The obtained modified MIT bag model alongside with the Friedmann equations are utilized to study the thermodynamical quantities in terms of the evolution time. The effect of GUP is set by inserting the parameter $\alpha = 10^{-5}\hspace{0.03cm}GeV^{-1}$. It is found that, the GUP effect on the evolution time of the universe makes it shorter, i.e the time span of each phase takes shorter time to transfer to another phase than their corresponding ones without GUP effect.}

\makeatletter
\gdef\@fpheader{}
\makeatother
\begin{document}
\maketitle
\flushbottom
\section{Introduction}\parindent 10mm
\hspace{0.9cm} It is widely known that an incompatibility between Quantum Mechanics and General Relativity exists. It has been suggested that the standard commutation relation, i.e. the Heisenberg Uncertainty Principle (HUP), would be considered for including gravitational effects \cite{Synder47,Yang47,Mead64,Karoly66}. Thus, HUP has been generalized to what is known as Generalized Uncertainty Principle (GUP).\par There are many versions of GUP which are distinguished by a GUP-parameter, which can be determined either theoretically \cite{Amati87,Gross87,Amati89,Konishi90,Veneziano90,Capozziello99,Maggiore93,Kempf95,Scardigli99,Adler99,Scardigli03} or phenomenological \cite{Das08,Das10,Das09,Farag1,Farag2}. It has been indicated that the equation of state (EoS) of an ideal Quark Gluon Plasma (QGP) is modified, when the GUP is applied, as a consequence of the modifications occurred to the corresponding thermodynamical quantities \cite{Farag3,Mashed14,Salem15}. The speed of sound and the entropy density was computed after considering GUP effect on EoS of an Ideal QGP \cite{Nasser18}.

It is believed that, the universe passed different phases during the history of its evolution from the moment of the big bang till now. After few microseconds of the big bang, the universe was in a deconfined form of matter (quarks and gluons) at very high temperature which is called quark gluon plasma (QGP) phase. As a consequence of expansion with cooling down, the hadronic phase of matter formed which interacts strongly. Quantum Chromodynamics (QCD) is the theory of the strong interactions which anticipates
two forms of matter at high and low temperature. The phase transition between both phases occurred at about $10^{-5}$ seconds
after the big bang \cite{Boeckel_11}.
In the recent years, and due to the developing in the experiments of heavy ion collisions and the lattice results, all the developments motivated the scientists to study in more depth the phase transition of matter. Exploration of the phase transition of the universe has been studied through different work. The evolution of the universe has been studied using the entropy of a hadrons and QGP systems by constructing a transport model \cite{Yunfei19}. It was showed that the crossover through entropy calculation is a good signal for QGP formation \cite{Yunfei19}. Various recent experiments insure different signatures of QGP formations such as, LHC \cite{Rafelski20}, STAR \cite{STAR2020}, ALICE \cite{ALICE14,ALICE19}, and CMS \cite{CMS11}. Additionally, the experimental and theoretical challenges for the QGP are discussed in detail \cite{Adam05}. For a review of the discovery of QGP see reference \cite{Rafelski17}.\par
  The Friedmann equations obviously describe the universe evolution through thermodynamics which employed in MIT bag model \cite{Chodos74, Johnson75}, i.e. with EoS \cite{Yagi05}. The realistic QCD EoS with bag EoS was studied in \cite{Florkowski11}, they showed a noticeable difference between both EoS in the energy density. Also, the bag model has showed a reasonable description of the phase transition, but there are other factors of the QGP should be theorized, e.g. the number of leptons and the strong magnetic field \cite{Sanches15,Fogaca15}. Since the universe started to expand from the moment of the big bang till the present time, and this expansion is accelerating through observations of distant supernovae \cite{Reiss1,Reiss2,Reiss3}. The importance of this study motivated the authors of the present work to figure out that period of time among different phases. We focus on the study of the QGP phase of matter and QCD matter passing through mixed phase between them. We investigate the effect of GUP on the evolution of the universe based on the Friedmann equations and the bag model modified with GUP.

Our paper is organized as follows. In Sec. \ref{sec:GUP}, we introduce GUP model and its impact on the different eras. Sec. \ref{sec:frame} contains two parts subsec. \ref{sec:Friedman}, and subsec.\ref{sec:effectGUP} introduce Friedmann equation used for describing the history of Universe, and the effect of GUP on it. Sec. \ref{sec:Time_temp} is devoted to show GUP effect on the bag equation of state with Friedmann equations.  Finally, in Sec.\ref{sec:results}, our results are presented and discussed.

\section{Generalized Uncertainty Principle (GUP)}\label{sec:GUP}
 \hspace{0.9cm} It is surmised that, a modification of the standard commutation relations would be exist at short distances. One of the most interesting forms of GUP has been proposed \cite{Farag1,Farag2} which is consistent with the string theory, doubly special relativity (DSR) theorem and black hole physics.
It suggests a minimal measurable length and a maximum observable momentum, GUP ensures that $[x_{i},x_{j}]= [p_{i},p_{j}]= 0$ (via the Jacobi identity):
\begin{equation}\label{farag gup}
    [x_{i},p_{j}]= i \hspace{0.02cm}\hbar[\delta_{ij}- \alpha(p\delta_{ij}+ \dfrac{p_{i}p_{j}}{p})+ \alpha^{2}(p^{2}\delta_{ij}+ 3 p_{i} p_{j})]
\end{equation}
where $\alpha= \alpha_{0}/M_{p}c= \alpha_{0} \ell_{p}/\hbar$ and $M_{p}c^{2}$ is Planck energy. $\ell_{p}$ and $M_{p}$ are Planck length and mass, respectively. $\alpha_{0}$ specifies the upper and lower bound on $\alpha$.\\
This GUP implies the presence of minimal measurable length and a maximum observable momentum given by
\begin{eqnarray}
  \Delta x_{min} &\approx& \alpha_{0} \ell_{p}  \\
  \Delta p_{max}  &=& M_{p}c/\alpha.
\end{eqnarray}
where $\Delta x\geq \Delta x_{min}$ and $\Delta p \leq \Delta p_{max}$. Hence, the generalized momentum for a particle which its energy is comparable with the Planck energy would be \cite{Farag1,Farag2}
\begin{equation}\label{genmom}
    p_{i}= p_{0i}(1- \alpha p_{0}+ 2\alpha^{2}p_{0}^{2})
\end{equation}
where $x_{i}= x_{0i}$ and $p_{0j}$ fulfill the canonical relations $[x_{0i}, p_{0j}]=\imath \hspace{0.03cm}\hbar\delta_{ij}$. Here, $p_{i}$ is interpreted as the momentum at high energies and $p_{0i}$ at low energies. This proposed GUP has an important suggestion that the space is discrete into fundamental measurable and minimum lengths which can be in Planck length scale. The space quantization has been presented within loop quantum gravity context \cite{loop1}.\\
The GUP effects have been studied on condensed, atomic matter, black holes at LHC \cite{cond1, atom1,bhlhc1}, the Liouville theorem (LT), and the weak equivalence principle (WEP) \cite{WEPLT}. Also, the small observed violations of the WEP in experiments of neutron interferometry can be explained using GUP \cite{VioWEP1, VioWEP2, VioWEP3}. \par It was found that the first bound for $\alpha_{0}$ is nearly $10^{17}$, which gives $\alpha\sim 10^{-2} \hspace{0.03cm} GeV^{-1}$. The other bound of $\alpha_{0}$ is nearly $10^{10}$, which gives $\alpha\sim 10^{-9} \hspace{0.03cm}GeV^{-1}$ \cite{atom1}. As stated in \cite{naser11}, the comparison with observations gives us the exact bound of $\alpha$. The implications of GUP effect can be directly measured in laboratories of quantum optics \cite{nature}.\par
There are another research scenario which try to construct a harmonious induced GUP due to gravity. This scenario is looking for an analog fulfillment of theoretical construction of High Energy like GUP originating from the minimal fundamental length of Dirac materials \cite{Iorio18,Iorio19,Iorio11} and the black holes in graphene \cite{Iorio13,Iorio12,Iorio14,Iorio15,Iorio20}.

\section{General Framework}\label{sec:frame}
\subsection{The Friedmann Equations}\label{sec:Friedman}
 \hspace{0.9cm}  In this part we discuss Friedmann equations which govern the expansion time of the universe. Additionally, it can be directly related to the equation of state, the essential part here is to study the energy density and pressure as a function of time. Moreover, the effect of GUP can be appeared in the equation of state in both phases (QGP, QCD) and mixed phase. Starting with Friedmann equation \cite{Yagi05} with zero cosmological constant.
\begin{equation}\label{eq:Friedmann1}
H^{2} = \left( \frac{\dot{a}}{a}\right)^{2} = \frac{8\pi G}{3} \varepsilon-\frac{K}{a^{2}}
\end{equation}
Where H, is the Hubble parameter in terms of the scale factor a(t), $\dot{a}$ is the derivative of the scale factor with respect to time, G is the gravitational constant and is taken as $(G = 1.2211\times 10^{19} \hspace{0.02cm} GeV )^{-2}$, $\varepsilon$ is the energy density and the parameter K represents the sign of the spatial curvature$:$ $K=-1$ (open space with negative curvature)$;$ $K=0$ (flat space)$;$ $K=+1$ (closed space with positive curvature). In our work $K=0$, since the universe is radiation-dominant in the early universe. Friedmann equation can be expressed in terms of energy density $\varepsilon$ and pressure $P$ as follows$:$
\begin{equation}\label{eq:Friedmann2}
\frac{\ddot{a}}{a} = -\frac{4\pi G}{3} \left( \varepsilon + 3P\right)
\end{equation}
The balance equation can be obtained from combining Eqs. (\ref{eq:Friedmann1}, \ref{eq:Friedmann2}) which yields \cite{Yagi05}
\begin{equation}\label{eq:balance}
\frac{d\varepsilon}{da} = -\frac{3}{a} \left( \varepsilon + P\right)
\end{equation}
The evolution of the universe within time is affected by different epochs. In the present work, we devote to study this effect in the QCD phase transition. So that, Eq.(\ref{eq:balance}) is re-written in order to describe the time history of the universe in QCD epoch based on the modified bag model, in other words with the effect of GUP on the construction of the bag model \cite{Mashed14}. Then, for flat space (i.e, K = 0), Eq.(\ref{eq:Friedmann1}) reads,
\begin{equation}\label{eq:balance1}
\frac{\dot{a}}{a} = -\frac{\dot{\varepsilon}}{3\left( \varepsilon + P\right) } = \sqrt{\frac{8\pi G}{3}\varepsilon}
\end{equation}
 Now, we turn to the construction of the equation of state for QGP and QCD phases based on Friedmann equations, linked with GUP effect.

 \subsection{GUP Effect on The Equation of State}\label{sec:effectGUP}
 The evolution epoch of matter had different phases, our main point here is studying these mentioned phases from the hot and dense phase of matter, which is characterized by QGP, mixed phase (intermediate), and the hadronization period which is characterized by QCD matter. Our main goal is to describe the bag model equation of states for both phases with the assumption of GUP modifications to the EoS for all phases.

 \begin{itemize}
 \item  \textbf{The Hadronic Phase} \\
  The modified energy density $ \varepsilon_{H}$ and the pressure $P_{H}$ in hadronic phase are given by \cite{Mashed14}

 \begin{equation}\label{eq:eos_hadronic}
 \begin{aligned}
  \varepsilon_{H} &= d_{H} \left( \frac{\pi^{2}}{30} T^{4}+ 3 \alpha_{1} T^{5}\right), \;\;\;\;\;
  &P_{H} &=  d_{H} \left( \frac{\pi^{2}}{90} T^{4}+  \alpha_{1}\hspace{0.01cm} T^{5}\right),
   \end{aligned}
  \end{equation}
  And the entropy in the hadronic phase in defined as \cite{Yagi05},
  \begin{equation}
   S_{H} = 4\left( \frac{P_{H}}{T}\right)
  \end{equation}
 The second term includes $\alpha$ which represents the modification of EoS due to GUP, where, $\alpha_{1} =\frac{24}{\pi^{2}}  \hspace{0.01cm}  \alpha \hspace{0.01cm} \zeta(5)$ with $\alpha$ is GUP parameter, and $\zeta(n)$ is the Riemann zeta function with $\zeta(5)=1.037$ and hadronic degree of freedom  $d_{H}$.
\item \textbf{QGP Phase} \\
 The modified energy density $\varepsilon_{QGP}$ and the pressure $P_{QGP}$ in QGP phase are given by \cite{Mashed14}

\begin{equation} \label{eq:eos_QGP}
\begin{aligned}
  \varepsilon_{QGP} &= d_{QGP} \left( \frac{\pi^{2}}{30} T^{4}+3 \alpha_{1} T^{5}\right)+B ,
  &P_{QGP} & =  d_{QGP} \left( \frac{\pi^{2}}{90} T^{4}+  \alpha_{1}\hspace{0.01cm} T^{5}\right)-B,
   \end{aligned}
  \end{equation}
Where $d_{QGP}$ is QGP degree of freedom, and B is the bag constant is modified according to the effect of GUP in the present work as,

\begin{equation}
B = d_{QGP}\hspace{0.03cm} T_{c}^{4}\hspace{0.03cm} \left( \frac{r-1}{r}\right) \hspace{0.03cm}\left( \frac{\pi^{2}}{90} + \alpha_{1} \hspace{0.03cm} T_{c}\right)
\end{equation}
Where r is ratio between QGP and hadronic degrees of freedom  $r = \frac{d_{QGP}}{d_{H}}$, and $T_{c} = 170 \hspace{0.05cm}MeV$ is the critical temperature between the two phases.
From Eq. (\ref{eq:eos_QGP}) the entropy of QGP can be defined as \cite{Yagi05};
\begin{equation}
S_{QGP} = \left(\frac{\varepsilon_{QGP}+ P_{QGP} }{T}\right)
\end{equation}

  Setting $\alpha = 0$, in Eqs. (\ref{eq:eos_hadronic}), and (\ref{eq:eos_QGP}), the bag model equations are recovered.

\item \textbf{The mixed phase}\\
 In the mixed phase, the energy density can be parametrized by a volume fraction factor as a function in time, f(t), as follows;
\begin{equation} \label {eq:mix_state}
\varepsilon(t) = \varepsilon_{H}(T_{c})\hspace{0.05cm} f(t) + \varepsilon_{QGP}(T_{c}) (1-f(t))
\end{equation}
 \end{itemize}
The three phases would be categorized according to the temperature and the time of evolution. Firstly, the QGP phase in which the temperature ($T > T_{c}$) and the time ($t < \tau_{c}$). Secondly, the mixed phase in which the temperature ($T = T_{c}$) and the time ($\tau_{c} < t < \tau_{H}$).
Where, $\tau_{c}$ and $\tau_{H}$ are the critical and starting time of hadronization of expansion process, respectively. Finally, the hadronic phase for temperature ($T < T_{c}$) and time ($t > t_{f}$).

\section{Time Evolution of The Universe Temperature}\label{sec:Time_temp}
\hspace{0.9cm} We elaborate the main assumption of GUP effect on the bag equation of state alongside with Friedmann equations. We use a simple steps to obtain the temperature of the different phases as a function of time:
\begin{itemize}
\item[\textbf{Case I}] QGP-phase for $T > T_{c}$ ($t < t_{c}$).\\
Using Eqs.(\ref{eq:eos_QGP}) and the second equality in Eq. (\ref{eq:balance1}), we obtained a differential equation as;

\begin{equation}\label{eq:T_L_Tc}
\frac{-\left[ \frac{2\pi^{2}}{15} + 15\hspace{0.02cm}\alpha_{1}\hspace{0.02cm} T\right] \hspace{0.03cm}dT}
{\left[\frac{2\pi^{2}}{15} T +12 \hspace{0.02cm} \alpha_{1}\hspace{0.02cm} T^{2}\right]
\sqrt{T^{4}\left( \sigma_{1}(T)+ \sigma_{2}(T)\hspace{0.03cm} T\right)} } = \frac{dt}{\lambda \sqrt{B}}
\end{equation}
Where $\lambda$ is the time scale and is defined as $\lambda = \sqrt{\frac{3}{8\hspace{0.02cm}\pi \hspace{0.02cm}G\hspace{0.02cm}B}}$,
 and
\begin{equation}
\sigma_{1}(T) = \frac{d_{QGP}\hspace{0.03cm}\pi^{2}}{30} \left[ 1 + \left( \frac{r-1}{3r}\right)\left( \frac{T_{c}}{T}\right)^{4}\right] ,  \nonumber
\\
\sigma_{2}(T) = 3 \hspace{0.02cm} \alpha_{1}\hspace{0.02cm} d_{QGP}\hspace{0.02cm} \left[ 1 + \left( \frac{r-1}{3r}\right)\left( \frac{T_{c}}{T}\right)^{5}\right]  \nonumber
\end{equation}
In order to obtain the temperature, we integrate with respect to time, then by using Newton-Raphson method to get the temperature as a function of time.
\item[\textbf{Case II}] Mixed phase for $T = T_{c}$ ($t_{c} < t < t_{f}$).\\
 Using Eq. (\ref{eq:mix_state}) and Eq. (\ref{eq:balance1}), one obtains
\begin{equation} \label{eq:T_eq_Tc}
\frac{\dot{a}}{a}=\frac{\dot{f}}{3\left[\frac{r}{r-1} -f \right] } = \frac{1}{\lambda} \sqrt{4(1-f)+ \frac{3}{r-1}}
\end{equation}
Therefore, Eq. (\ref{eq:T_eq_Tc}) is solved analytically \cite{Yagi05},
\begin{equation}
f(t) = 1- \frac{1}{4(r-1)}\left[ tan^{2}\left( \frac{3}{2} \frac{\lambda^{-1}}{\sqrt{r-1}} (t-t_{I}) - tan^{-1}\sqrt{4\hspace{0.02cm}r-1}\right)- 3 \right]
\end{equation}
 As a matter of fact, the mixed phase ends up at $f(t_{f}) = 1$, one obtains
 \begin{equation}
 t_{f}-t_{I} = \lambda \hspace{0.02cm} \frac{2\sqrt{r-1}}{3}\left[ tan^{-1}\sqrt{4\hspace{0.02cm}r-1} - tan^{-1}\sqrt{3}\right]
 \end{equation}
 \item[\textbf{Case III}] Hadronic-phase for $T < T_{c}$ ($ t > t_{f}$)\\
 In this case the temperature decreases to be less than the critical one, in other words the matter freezes during the expansion to be QCD matter. Using Eq.(\ref{eq:eos_hadronic}) into Eq. (\ref{eq:balance1}), we have a differential equation as;

 \begin{equation}
 \frac{-d\varepsilon_{H}}{4\hspace{0.02cm}\varepsilon_{H} \sqrt{\varepsilon_{H}}} =  \frac{dt}{\lambda \sqrt{B}}
 \end{equation}
 Then, it can be solved analytically as follows,
 \begin{eqnarray}
 t-t_{f} = \lambda \sqrt{\frac{15\hspace{0.02cm}B}{2}}
\left(  \frac{1}{ T^{2} \sqrt{d_{H}(\pi^{2}+30\hspace{0.02cm}\alpha_{1}\hspace{0.02cm} T) }}-  \frac{1}{ T_{c}^{2} \sqrt{d_{H}(\pi^{2}+30\hspace{0.02cm}\alpha_{1}\hspace{0.02cm} T_{c}) }}\right)
  \end{eqnarray}
\end{itemize}
\section{Results and Discussion}\label{sec:results}
\hspace{0.9cm} In this section, we shall perform a numerical analysis and investigate the effect of GUP on the temperature of the universe during the evolution time. The expansion of the universe comprises various eras and different temperature from the QGP phase, through a mixed phase, to the hadronic phase.

\begin{figure}[!htb]
\includegraphics[width=14.cm, height=8.cm]{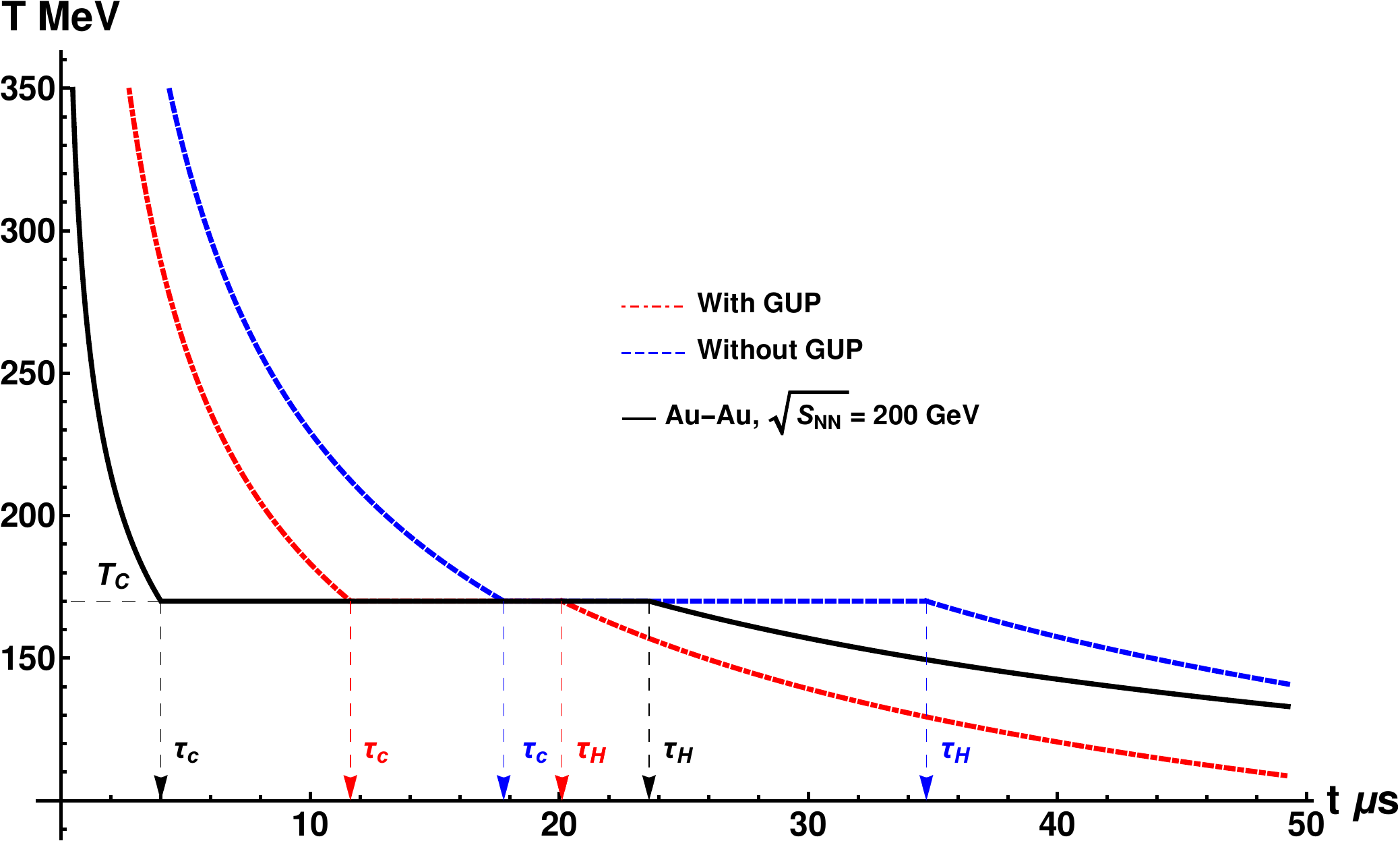}
 \centering
 \caption[figtopcap]{ The calculated temperature as a function of time with the effect of GUP in the present work. \label{fig:comparison}Temperature as a function of time for QCD phase transition, solid black curve at $T>T_{c}$, online dashed blue line is the temperature in the mixed phase i.e at $T = T_{c}$, and online solid red line at $T < T_{c}$.}
\end{figure}

Figure (\ref{fig:comparison}) illustrates the relation between the temperature and the time of expansion of the matter from QGP phase to QCD phase in which the mixed phase occurs at a time between the two phases. There are three different plots are represented, the solid black for Au-Au collision \cite{Wang97}, the present model is the modified bag model due to GUP-effect is represented by dotted-dashed red, and finally the bag model \cite{Yagi05} is represented by the dashed blue.\par
 In the present work, we have selected the critical temperature $T_{c} =170 \hspace{0.03cm} MeV$, this figure is divided into three regions characterized by the time ($\tau_{c}, \tau_{H}$) which are the end time of QGP, and the end time of the mixed phase, respectively. In other words, the time of the mixed phase starting, and the time of the hadronic phase starting.

 It is worth to shed light on the limits that each model and$/$or experimental data exhibit in the time scale. These limits are represented by black, red, blue arrows. These limits are summarized in table (\ref{tab:time_evol}), one can notice that, the QGP time span is very short in the experimental data which ends at just 4 $\mu sec$, this makes the experimental attempts still do not confirm the observation of QGP phase. While the bag model and the GUP is rather longer.
  It is noticeable that, the mixed phase varies between the mentioned work above. In the present work, with GUP effect, the mixed phase is tiny compared with the corresponding experimental data and bag model. This means that, GUP effect shorten the mixed phase and the QGP-phase is roughly change rapidly to the hadronic phase. The life time of the mixed phase is also determined in table (\ref{tab:time_evol}) which is presented by the ratio of $\frac{\tau_{H}}{\tau_{c}}$. The three different eras can be illustrated as:

\begin{itemize}
\item  For $T>T_{c}$, ($ \tau < \tau_{c})$ \\
It is the period in which QGP phase dominates and at temperature ($T = 2T_{c}$) it begins to expand rapidly to reach at initial time $\tau_{c}$.
\item For $T = T_{c}$, ($ \tau_{c} < \tau < \tau_{H}$) \\
A mixture of QGP and QCD states can coexist between the two borders beginning at $\tau_{c}$ and ending at  $\tau_{H}$.
\item  For $T < T_{c}$, ($\tau > \tau_{H}$) \\
The last period and at longer time, the matter cools down to be in the hadronic phase.
\end{itemize}
\begin{table}[ht!]
\centering
\begin{tabular}{ c | c }
\hline
  &  $\frac{\tau_{H}}{\tau_{c}}$  \\
\hline
Au-Au collision, $\sqrt{S_{NN}} = 200 \hspace{0.05cm}GeV $ \cite{Wang97} &   5.9  \\
Present model "GUP" &    1.73 \\
MIT Bag model \cite{Yagi05} &  1.96\\
\hline
\end{tabular}\caption{\label{tab:time_evol} The critical time and hadronization time for different models.}
\end{table}
As a matter of fact, the temperature is determined as a function of time. Therefore, the thermodynamical quantities can be evaluated as a dynamical observables. In other words, the thermodynamical quantities are calculated as a function of time, the following figure (\ref{fig:thermodyn1}) illustrates the evolution of QGP thermodynamics pressure, energy density in $MeV^{4}$ versus the time.

 The upper panel of figure (\ref{fig:3a}) represents the pressure versus time for QGP phase from time $(t = 0-10)$ $\hspace{0.07cm} \mu sec$, the solid line represents the pressure with the effect of GUP and the dotted-dashed without the GUP effect. One notice that, both curves behave qualitatively as well, but the addition of GUP lowers the pressure than the other one, approximately half the value in particular time ($t\sim 1-8 \hspace{0.04cm} \mu sec $). It is worth to note that, at ($t\sim 7 \hspace{0.08cm} \mu sec$), the pressure tends to its minimum value while the other one (without GUP) still survives for longer time ($t > 10 \hspace{0.08cm}$).

 This might be attributed to that, GUP effect makes the phase transition takes shorter time to reach to another phase.
For the energy density fig.(\ref{fig:3b}), has the same feature for the transition to the next phase in shorter time with GUP effect than without the GUP effect, these results is comparable to the results in \cite{Rajagopal}.
\begin{figure}[!htb]
\subfigure[The quark-gluon plasma pressure versus the evolution time with the effect of GUP (solid-line), and without GUP (Dotted-dashed)]{\label{fig:3a} \includegraphics[width=12.cm, height=7.cm]{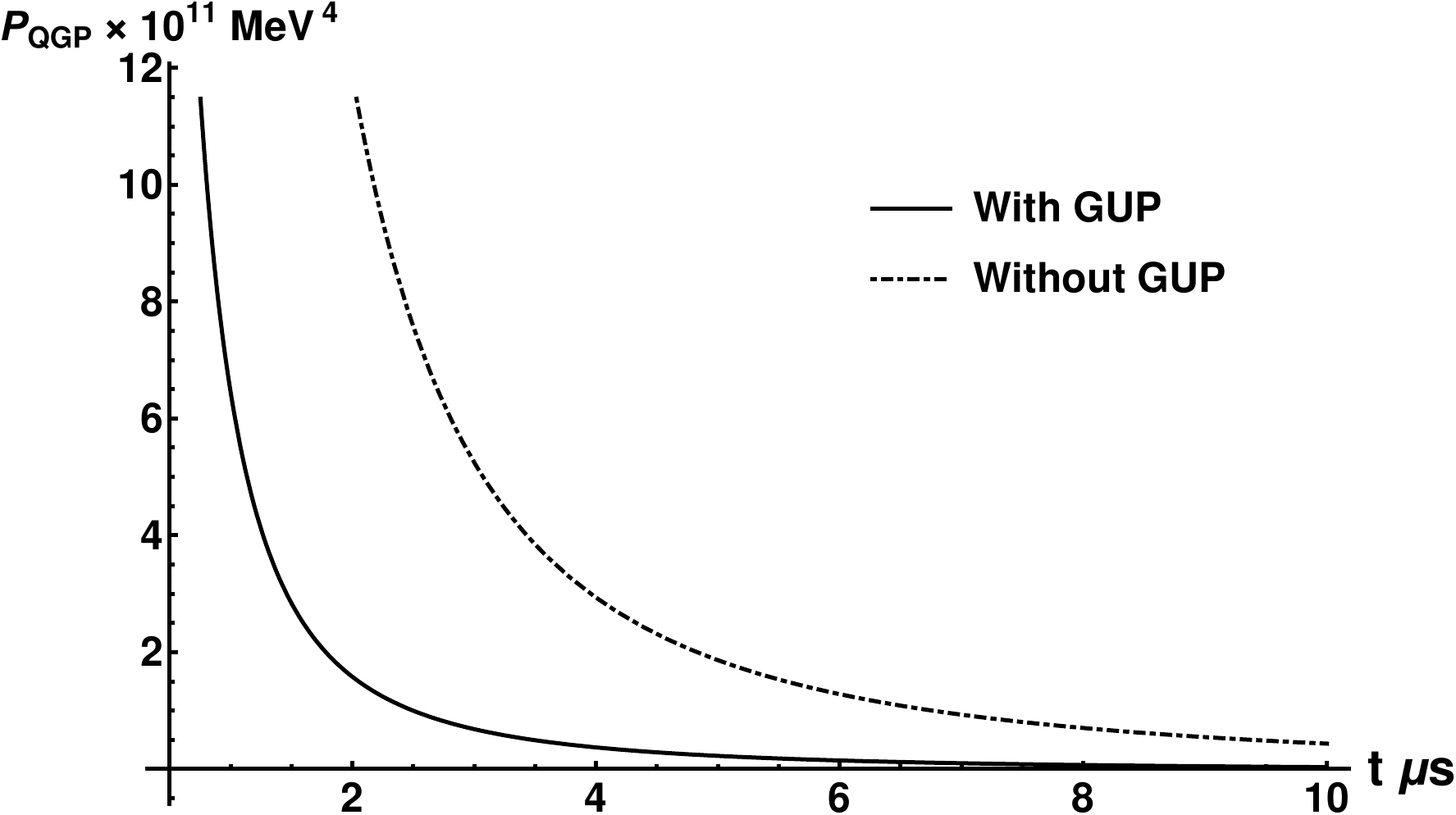}}
\subfigure[The quark-gluon plasma energy density versus the evolution time with the same two cases in \ref{fig:3a}]{\label{fig:3b} \includegraphics[width=12.cm,height=7.cm]{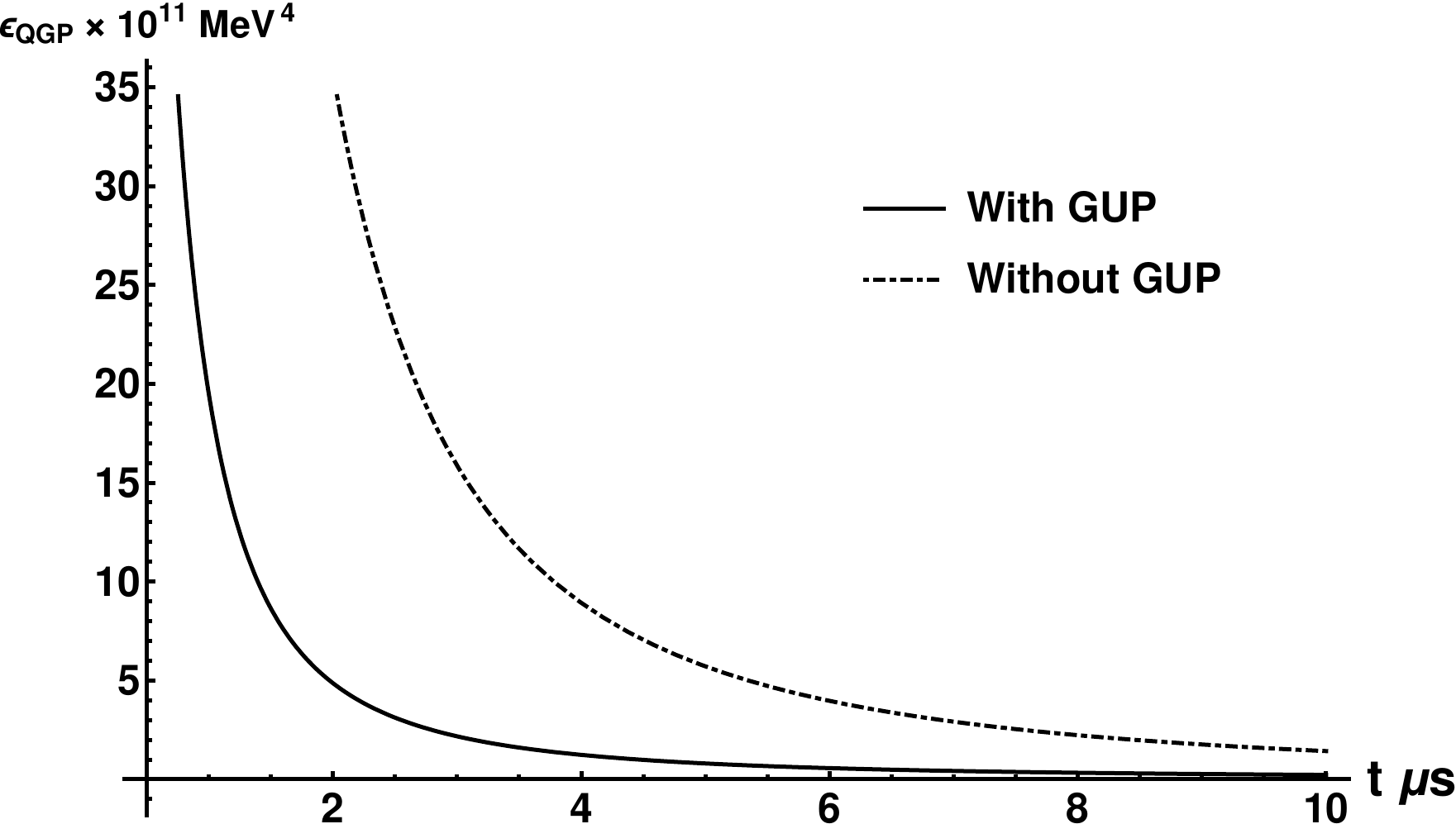}}
\caption[fittopcap]{The bag model equation of state pressure \subref{fig:3a} and energy density \subref{fig:3b} for the QGP with and without the effect of GUP (solid-Black) and (Dotted-Dashed) respectively}\label{fig:thermodyn1}
\end{figure}
\begin{figure}[!htb]
\includegraphics[width=12.cm,height=8.cm]{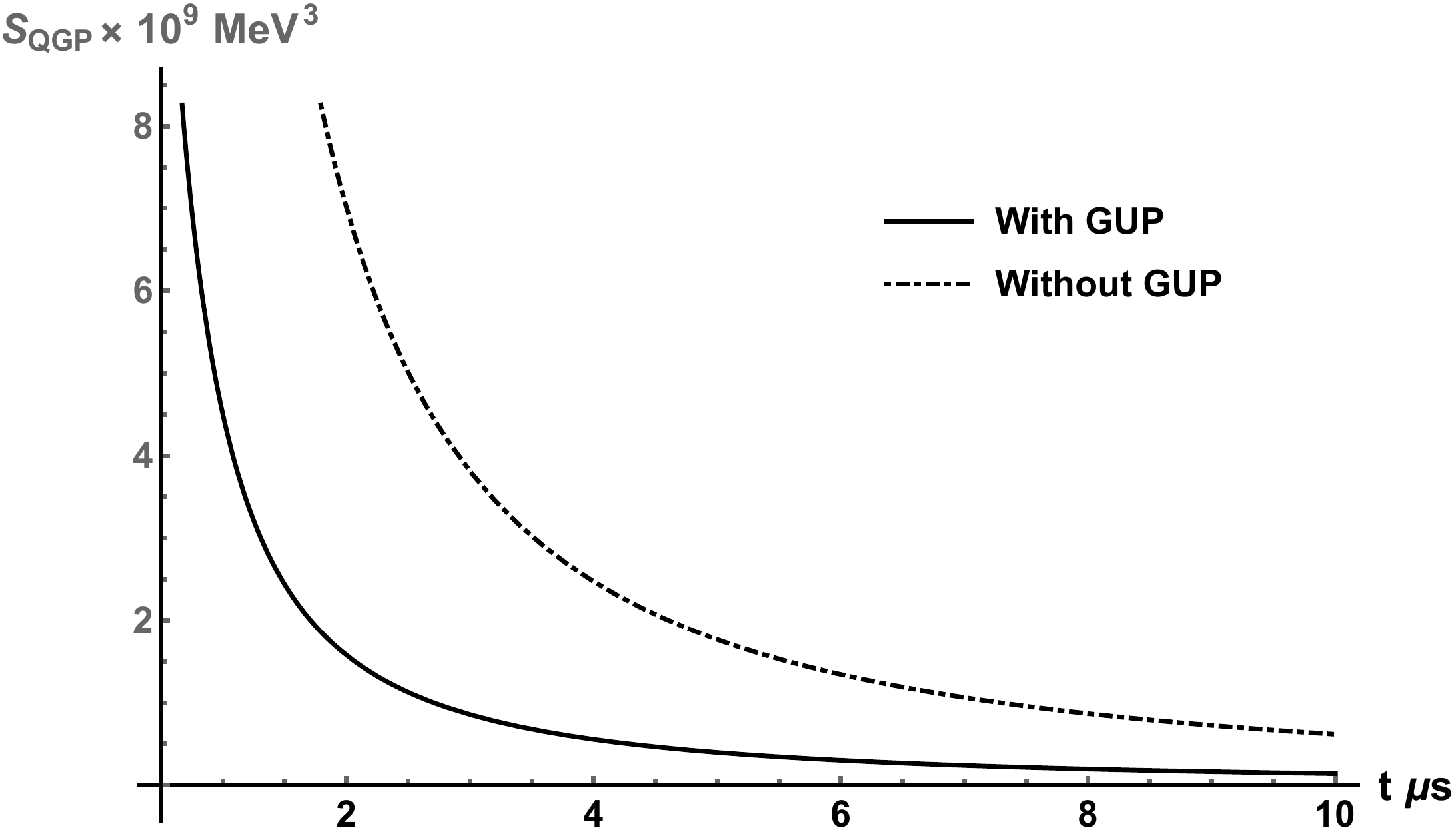}
\caption{The bag model entropy for the QGP with and without the effect of GUP (solid-Black) and (Dotted-Dashed) respectively}\label{fig:thermodyn11}
\end{figure}
Figure (\ref{fig:thermodyn11}) shows the entropy of QGP versus time and one can see the same notice as pressure and energy density. Finally, it is worth to shed light on the different quantities in figures (\ref{fig:thermodyn1}, \ref{fig:thermodyn11}), it is remarkable that at relatively small time ($0 < t \lesssim 5 \hspace{0.03cm}\mu s$), both the GUP effect and the original one are drifting away, and as the time becomes longer than $5 \hspace{0.03cm}\mu s$, they are approaching each other. This means that, the thermodynamical quantities in the presence of GUP effect can be compared with their corresponding ones in the absence of GUP effect at roughly larger time, i.e. lower temperature.
\section{Conclusion}
\hspace{0.9cm} In the present work, we have explored the effect of GUP on the evolution time of the universe in the context of the equation of state. We have obtained analytic expressions of the modified Friedman equations using GUP effect. We can also go some way towards modifying the first Friedmann equation with GUP contribution by establishing the modified equation of states of the MIT bag model in which the latter describes the energy density, the pressure and the entropy in hadronic and QGP phases. This can be done by setting thermodynamics into the picture. GUP effect appears through the modification in the uncertainty principle with an extra parameter $\alpha$ which discussed in details in Sec.\ref{sec:GUP}. In conclusion, the present work indicates that, GUP impacted the evolution time of the universe which made it shorter.

\acknowledgments
This research is supported by Academy of Scientific Research and Technology (ASRT), Egypt. (\textbf{ScienceUp $-$ Grants (2020)  No.6720}) granted to the faculty of science, Benha University.
\newpage
\providecommand{\href}[2]{#2}
\begingroup\raggedright

\endgroup
\end{document}